\documentclass[12pt]{article}
\usepackage{amsmath} 
\usepackage{cite}
\input{epsf} 
\newcommand\as{\alpha_{\mathrm{S}}} 
\newcommand\f[2]{\frac{#1}{#2}} 
\def\beq{\begin{equation}} 
\def\eeq{\end{equation}} 
\def\beeq{\begin{eqnarray}} 
\def\eeeq{\end{eqnarray}} 
 
\def\to{\rightarrow}
 
\def\nn{\nonumber}

\def\ms{${\overline {\rm MS}}$}

\def\ep{\epsilon}

\begin{document} 
\begin{titlepage}
\begin{flushright}
ZU-TH 20/14
\end{flushright}
\renewcommand{\thefootnote}{\fnsymbol{footnote}}
\vspace*{1cm}

\begin{center}
{\Large \bf Threshold resummation at N$^3$LL accuracy\\[0.3cm]
and soft-virtual cross sections at N$^3$LO}
\end{center}

\par \vspace{2mm}
\begin{center}
{\bf Stefano Catani${}^{(a)},$ Leandro Cieri${}^{(b)},$
Daniel de Florian${}^{(c)}$, Giancarlo Ferrera${}^{(d)},$\\
}
and
{\bf Massimiliano Grazzini${}^{(e)}$\footnote{On leave of absence from INFN, Sezione di Firenze, Sesto Fiorentino, Florence, Italy.}}\\

\vspace{5mm}

$^{(a)}$ INFN, Sezione di Firenze and Dipartimento di Fisica e Astronomia,\\ 
Universit\`a di Firenze,
I-50019 Sesto Fiorentino, Florence, Italy\\

$^{(b)}$ Dipartimento di Fisica, Universit\`a di Roma ``La Sapienza'' and\\
INFN, Sezione di Roma, I-00185 Rome, Italy\\

${}^{(c)}$ Departamento de F\'\i sica, FCEYN, Universidad de Buenos Aires,\\
(1428) Pabell\'on 1 Ciudad Universitaria, Capital Federal, Argentina\\

${}^{(d)}$ Dipartimento di Fisica, Universit\`a di Milano and\\ INFN, Sezione di Milano,
I-20133 Milan, Italy\\

$^{(e)}$ Physik-Institut, Universit\"at Z\"urich, CH-8057 Z\"urich, Switzerland

\vspace{5mm}

\end{center}

\par \vspace{2mm}
\begin{center} {\large \bf Abstract} \end{center}
\begin{quote}
\pretolerance 10000

We consider QCD radiative corrections to the production of
colourless high-mass systems in hadron collisions.
We show that the recent computation of the soft-virtual corrections
to Higgs boson production at N$^3$LO \cite{Anastasiou:2014vaa} together with the universality structure
of soft-gluon emission
can be exploited to extract the general expression of
the hard-virtual coefficient that contributes to threshold resummation
at N$^3$LL accuracy.
The hard-virtual coefficient is directly related to the process-dependent
virtual amplitude through a universal (process-independent)
factorization formula that we explicitly evaluate up to
three-loop order.
As an application, we present the explicit expression
of the soft-virtual N$^3$LO corrections for the production of
an arbitrary colourless system. In the case of the Drell--Yan process, we confirm the recent
result of Ref.~\cite{Ahmed:2014cla}.

\end{quote}

\vspace*{\fill}
\begin{flushleft}
May 2014

\end{flushleft}
\end{titlepage}

\newpage
\setcounter{page}{1}
\setcounter{footnote}{1}
\renewcommand{\thefootnote}{\fnsymbol{footnote}}

The authors of Ref.~\cite{Anastasiou:2014vaa} have recently presented the result
of the calculation of the cross section for the threshold production of the Higgs
boson at hadron colliders at the next-to-next-to-next-to-leading order 
(N$^3$LO) in perturbative QCD. This result has prompted the observation 
\cite{Ahmed:2014cla} that the Higgs boson calculation contains information 
on soft-gluon radiation that
can be implemented to explicitly determine the N$^3$LO threshold cross section 
for the Drell--Yan (DY) process. In the present contribution, we exploit the
universality (process-independent) structure \cite{Catani:2013tia} 
of soft-gluon contributions near partonic threshold and the specific calculation of
Ref.~\cite{Anastasiou:2014vaa}. We show how the results of 
Refs.~\cite{Anastasiou:2014vaa} and \cite{Catani:2013tia} can be 
straightforwardly
combined and used
to extract the general expression of the hard-virtual coefficient that contributes
to threshold resummation at next-to-next-to-next-to-leading-logarithmic 
(N$^3$LL) accuracy for the cross section of a generic (and arbitrary)
colourless high-mass system produced in hadron collisions.
The threshold resummation formula for the production cross section can also be 
perturbatively expanded up to N$^3$LO, and for the specific case of the DY 
process we recover the result of Ref.~\cite{Ahmed:2014cla}.

The N$^3$LO Higgs boson results of Ref.~\cite{Anastasiou:2014vaa} 
complete a cross section
calculation that requires the evaluation of several independent ingredients
related to collinear-counterterm factors \cite{Moch:2004pa, counterterms}
and to real- \cite{real, Li:2013lsa} 
and virtual-radiation \cite{uvren, Chetyrkin:1997un, qgformfactor}
contributions. One of these ingredients is the three-loop virtual amplitude
\cite{Chetyrkin:1997un, qgformfactor} $gg \to H$ for Higgs boson production
through gluon fusion
(the three-loop results of Refs.~\cite{Chetyrkin:1997un, qgformfactor}
use the large-$m_{top}$ approximation).
As discussed in Ref.~\cite{Catani:2013tia}, all-order soft-gluon resummation
\cite{Sterman:1986aj, Catani:1989ne, Catani:1990rp}
for the hadroproduction cross section of a generic colourless high-mass system
can be expressed in a process-independent form, whose sole process-dependent
information is encoded in the virtual amplitude of the specific process.
Therefore, using the cross section of Ref.~\cite{Anastasiou:2014vaa} and the
virtual amplitude of Refs.~\cite{Chetyrkin:1997un, qgformfactor} for the
specific case of
Higgs boson production, we can apply the formulation of 
Ref.~\cite{Catani:2013tia} and we can explicitly determine the entire 
process-independent information that contributes to soft-gluon resummation
for a generic production process up to the three-loop level.
In the following we recall the formalism of soft-gluon resummation 
(by mainly following the notation of Sect.~5 in Ref.~\cite{Catani:2013tia}) 
and we present and illustrate our three-loop results.

We consider the inclusive hard-scattering reaction
\begin{equation}
h_1(p_1)+h_2(p_2)\to F(\{q_i\})+X\, ,
\label{class}
\end{equation}
where the collision of the two hadrons $h_1$ and $h_2$ with momenta $p_1$ and
$p_2$ produces the triggered final state $F$, and $X$ denotes the accompanying
final-state radiation. 
The observed final state $F$ is a {\em generic} system of one or more
{\em colourless} particles (with momenta $q_i$),
such as lepton pairs (produced by the DY mechanism), photon pairs, 
vector bosons, Higgs boson(s), and so forth.
We focus on the {\em total}
cross section\footnote{The formalism of soft-gluon resummation can be further
elaborated and extended to include the dependence on final-state kinematical 
variables such as, for instance, the rapidity of the final state $F$ (see, e.g.,
Refs.~\cite{Catani:1990rp, Mukherjee:2006uu, Ravindran:2006bu, Ahmed:2014uya}).} 
for the process in Eq.~(\ref{class}) at fixed value $M$ of the
invariant mass of the triggered final state $F$ (i.e., we integrate the
differential cross section over the momenta $q_i$ with the constraint
$(\sum_i q_i)^2=M^2$).
In the simplest case, the final-state system $F$ consists of a
single (`on-shell') particle of mass $M$ (for example, $F$ can be a vector boson
or a Higgs boson). 
The total cross section $\sigma_F(p_1,p_2;M^2)$
for the production of the system $F$ is computable in QCD perturbation theory
according to the following factorization formula:
\begin{equation}
\sigma_F(p_1,p_2;M^2) = \sum_{a_1, a_2} \int_0^1 dz_1 \int_0^1 dz_2 \;
\;\hat{\sigma}^F_{a_1 a_2}({\hat s}=z_1z_2s;M^2; \as(M^2)) 
\; f_{a_1/h_1}(z_1,M^2)  \; f_{a_2/h_2}(z_2,M^2) 
\;\;,
\label{sigtot}
\end{equation}
where $s=(p_1+p_2)^2 \simeq 2p_1\cdot p_2$,
$\hat{\sigma}^F_{a_1 a_2}$ is the total partonic cross section for the
inclusive partonic process $a_1 a_2 \to F + X$ and, for simplicity, the parton
densities $f_{a_i/h_i}(z_i,M^2)$ $(i=1,2)$ are evaluated at the scale $M^2$
(the inclusion of an arbitrary factorization scale $\mu_F$ in the parton
densities and in the partonic cross sections can be implemented in a
straightforward way by using the Altarelli--Parisi evolution equations of
$f_{a/h}(z,\mu_F^2)$).
The partonic cross section
$\hat{\sigma}^F_{a_1 a_2}({\hat s};M^2; \as(M^2))$ depends on the mass $M$
of the system $F$, on the centre--of--mass energy $\sqrt {\hat s}$ of 
the colliding partons $a_1$ and $a_2$, 
and it is a renormalization-group invariant quantity
that can be perturbatively computed as series expansion in powers of $\as(M^2)$.
Considering, for instance, the inclusive partonic channel $c {\bar c} \to F + X$,
we can write
\begin{equation}
\label{sigmahat}
\hat{\sigma}^F_{c{\bar c}}({\hat s};M^2; \as(M^2))= 
\sigma^{(0)}_{c{\bar c}\to F}(M^2;
\as(M^2))\sum_{n=0}^{\infty}\left(\f{\as(M^2)}{\pi}\right)^n\,z
\;g^{F(n)}_{c{\bar c}}(z)
\end{equation}
where $z=M^2/{\hat s}$,
\begin{equation}
g_{c{\bar c}}^{F(0)}(z)=\,\delta(1-z) \;\;,
\end{equation}
and $\sigma^{(0)}_{c{\bar c}\to F}$ is the lowest-order cross section
for the partonic process $c{\bar c}\to F$.
Since the system $F$ is colourless, the lowest-order cross section is determined by the
partonic processes of quark--antiquark annihilation ($c=q,{\bar q}$)
and/or gluon fusion ($c=g$) (in the case of $q{\bar q}$-annihilation
the quark and antiquark can have different flavours, such as, for instance,
if $F=W^{\pm}$).
Perturbative expressions that are analogous to Eq.~(\ref{sigmahat}) can be 
written for the partonic cross sections $\hat{\sigma}^F_{a_1 a_2}$
of all the other partonic channels.
Using the renormalization-group evolution of the QCD running coupling 
$\as(q^2)$, 
we can equivalently expand $\hat{\sigma}^F_{a_1 a_2}$ in powers of
$\as(\mu_R^2)$, with corresponding perturbative coefficients $g_{a_1 a_2}^{F(n)}$
that explicitly depend on $M^2/\mu_R^2$, where $\mu_R$ is an arbitrary
renormalization scale. Throughout the paper we use parton densities as defined
in the \ms\ factorization scheme, and $\as(q^2)$ is the QCD running coupling in
the \ms\ renormalization scheme.

The kinematical variable $z=M^2/{\hat s}$ in Eq.~(\ref{sigmahat}) 
parametrizes the distance from the 
partonic threshold. The limit $z \to 1$ specifies the 
kinematical region that is close to the partonic threshold.
In this region 
the partonic cross section 
$\hat{\sigma}^F_{a_1 a_2}$ receives large QCD radiative corrections 
that are proportional to the singular functions
\begin{equation}
\label{plusdist}
{\cal D}_m(z) \equiv \left[\f{1}{1-z} \ln^m(1-z)\right]_+ \;\;, \;\;\; \quad
(m=0,1,\dots) \;,
\end{equation}
where the subscript `$+$' denotes the customary
`plus-distribution'. The all-order perturbative 
resummation of these logarithmic contributions (including all the singular
contributions that are proportional to $\delta(1-z))$)
can be systematically performed by working 
in Mellin ($N$-moment) space
\cite{Sterman:1986aj, Catani:1989ne}. The Mellin transform 
$\hat{\sigma}_{N}(M^2)$ of the partonic cross section 
$\hat{\sigma}({\hat s};M^2)$ is defined as
\begin{equation}
\hat{\sigma}^F_{a_1 a_2, \,N}(M^2; \as(M^2)) \equiv
\int_0^1 dz \; z^{N-1} \;\hat{\sigma}^F_{a_1 a_2}({\hat s}=M^2/z;M^2; \as(M^2))
\;\;.
\end{equation}
In Mellin space, the threshold region $z \to 1$ corresponds to the limit
$N \to \infty$, and the plus-distributions of Eq.~(\ref{plusdist}) 
become powers of $\ln N$
( $\left(\f{1}{1-z} \ln^m(1-z)\right)_+ \to \ln^{m+1} N + `{\rm subleading \;
logs}'$). These logarithmic contributions are evaluated to all perturbative 
orders by using threshold resummation \cite{Sterman:1986aj, Catani:1989ne}.
Neglecting terms that are relatively suppressed by powers of $1/N$ in the
limit $N \to \infty$, we write
\begin{equation}
\hat{\sigma}^F_{c {\bar c}, \,N}(M^2; \as(M^2)) =
\hat{\sigma}^{F ({\rm res})}_{c {\bar c}, \,N}(M^2; \as(M^2))
\; \Bigl[ 1 + {\cal O}(1/N) \Bigr] \;\;.
\label{thresfor}
\end{equation}
Note that we are considering only the partonic channel $\,c{\bar c} \to F +X$,
with $c{\bar c}=q{\bar q}$ and/or $c{\bar c}=gg$, since the other partonic channels
give contributions that are of ${\cal O}(1/N)$.
In this paper, we use the Mellin-space formalism of threshold resummation
\cite{Sterman:1986aj, Catani:1989ne}
that we have just introduced. Related formulations of threshold resummation
for hadron--hadron collisions
can be found, for instance, in Ref.~\cite{Ravindran:2005vv} (which is exploited to 
derive the results of Ref.~\cite{Ahmed:2014cla}) 
and in Refs.~\cite{Forte:2002ni, Idilbi:2006dg, Becher:2007ty}.

The expression $\hat{\sigma}^{F ({\rm res})}_{c {\bar c}, \,N}$ in 
the right-hand
side of Eq.~(\ref{thresfor}) embodies all the perturbative terms that are
logarithmically enhanced or {\em constant} in the limit $N \to \infty$.
The partonic cross section $\hat{\sigma}^{F ({\rm res})}_{c {\bar c}, \,N}$
has a universal (process-independent)  all-order structure that is given by the
following threshold-resummation formula
\cite{Sterman:1986aj, Catani:1989ne, Catani:1990rp, Vogt:2000ci,
Catani:2003zt, Moch:2005ba}:
\begin{equation}
\hat{\sigma}^{F ({\rm res})}_{c {\bar c}, \,N}(M^2; \as(M^2))
= \sigma^{(0)}_{c{\bar c}\to F}(M^2; \as(M^2)) \;
C^{\, \rm th}_{c{\bar c}\to F}(\as(M^2)) \; \Delta_{c, \,N}(M^2) \;\;.
\label{thallorder}
\end{equation}
The factor $\sigma^{(0)}_{c{\bar c}\to F}$ obviously depends on the produced 
final-state system $F$, and it is simply proportional to the square of the
lowest-order scattering amplitude ${\cal M}_{c{\bar c}\to F}^{(0)}$
(see Eq.~(\ref{ampli}))
of the partonic process $c{\bar c}\to F$.
The factor $C^{\, \rm th}_{c{\bar c}\to F}$ also depends 
on the produced final-state system $F$ and, therefore, it includes a 
process-dependent component.
The factor $\Delta_{c, \,N}$ is process-independent: it does not
depend on the final-state system $F$, and it only 
depends on the type ($c=q$ or $c=g$) of colliding partons.

The factor $\Delta_{c, \,N}$ is entirely due to soft-parton radiation
\cite{Sterman:1986aj, Catani:1989ne}.
This radiative factor
resums all the perturbative contributions $\as^n \ln^m N$
(including some constant terms, i.e. terms with $m=0$), and it has
the following all-order form: 
\begin{equation}
\label{delformfact}
\Delta_{c, \,N}(M^2)= \exp \left\{ \int_{0}^{1} dz \,\frac{z^{N-1}-1}{1-z} 
\left[ 2 \int_{M^2}^{(1-z)^2M^2} \frac{dq^2}{q^2} 
 A_c(\as(q^2))  + D_c(\as((1-z)^2M^2)) 
 \right] \right\} 
\;\;,
\end{equation}
where $A_c(\as)$ and $D_c(\as)$ are perturbative series in $\as$,
\begin{align}
\label{afun}
A_c(\as)&=\left(\frac{\as}{\pi}\right) A^{(1)}_c
+\left(\frac{\as}{\pi}\right)^2 A^{(2)}_c +
\left(\frac{\as}{\pi}\right)^3 A^{(3)}_c
+\left(\frac{\as}{\pi}\right)^4 A^{(4)}_c
+{\cal O}(\as^5) \;,\\
D_c(\as)&=\left(\frac{\as}{\pi}\right)^2 D^{(2)}_c +
\left(\frac{\as}{\pi}\right)^3 D^{(3)}_c+{\cal O}(\as^4) \;.
\label{dfun}
\end{align}
The function $A_c(\as)$ is produced by radiation that is soft and collinear
to the direction of the colliding partons $c$ and ${\bar c}$. The effect
of soft non-collinear radiation is embodied in the function $D_c(\as)$.
The perturbative coefficients 
$A^{(1)}_c, A^{(2)}_c$
\cite{Catani:1989ne, Catani:1990rr, Catani:1998tm} 
and 
$A^{(3)}_c$
\cite{Moch:2004pa, Moch:2005ba} are explicitly known. They read
\begin{align}
A_c^{(1)}&=C_c \;\;,\nn\\
A_c^{(2)}&=\f{1}{2}K\,C_c~~,~~~~~K=C_A\left(\f{67}{18}-\f{\pi^2}{6}\right)-\f{5}{9}
n_F \;,\nn\\
\label{Acoef}
A_c^{(3)}&=C_c\Bigg(\left(\f{245}{96}-\f{67}{216}\pi^2+\f{11}{720}\pi^4+\f{11}{24}\zeta_3\right)C_A^2+\left(-\f{209}{432}+\f{5}{108}\pi^2-\f{7}{12}\zeta_3\right)C_A\, n_F\nn\\
&+\left(-\f{55}{96}+\f{1}{2}\zeta_3\right)C_F\, n_F-\f{1}{108}n_F^2\Bigg) \;,
\end{align}
where $n_F$ is the number of quark flavours, $N_c$ is the number of colours, and
the colour factors are $C_F=(N_c^2-1)/(2N_c)$ and $C_A=N_c$ in $SU(N_c)$ QCD.
The colour coefficient $C_c$ depends on the type $c$ of colliding partons,
and we have $C_c=C_F$ if $c=q$ and $C_c=C_A$ if $c=g$.
The perturbative expansion of $D_c(\as)$ starts at ${\cal O}(\as^2)$
(i.e., $D^{(1)}_c=0$), and the perturbative coefficients $D^{(2)}_c$ 
\cite{Vogt:2000ci, Catani:2001ic}
and $D^{(3)}_c$ \cite{Moch:2005ky, Laenen:2005uz} 
are explicitly known. They read
\begin{align}
D_c^{(2)}&=C_c\left(C_A\left(-\f{101}{27}+\f{11}{18}\pi^2+\f{7}{2}\zeta_3\right)+n_F\left(\f{14}{27}-\f{1}{9}\pi^2\right)\right)
\;,\nn\\
\label{Dcoef}
D_c^{(3)}&=C_c\Bigg(C_A^2\left(-\f{297029}{23328}+\f{6139}{1944}\pi^2-\f{187}{2160}\pi^4+\f{2509}{108}\zeta_3-\f{11}{36}\pi^2\zeta_3-6\zeta_5\right)\nn\\
&+C_A\, n_F\left(\f{31313}{11664}-\f{1837}{1944}\pi^2+\f{23}{1080}\pi^4-\f{155}{36}\zeta_3\right)\nn\\
&+C_F\,
n_F\left(\f{1711}{864}-\f{1}{12}\pi^2-\f{1}{180}\pi^4-\f{19}{18}\zeta_3\right)+n_F^2\left(-\f{58}{729}+\f{5}{81}\pi^2+\f{5}{27}\zeta_3\right)\Bigg)
\;.
\end{align}
Using Eq.~(\ref{delformfact}), the coefficients 
$A^{(1)}_c, A^{(2)}_c, A^{(3)}_c, D_c^{(2)}, D_c^{(3)}$ in
Eqs.~(\ref{Acoef})--(\ref{Dcoef})
and the coefficient $A^{(4)}_c$ in Eq.~(\ref{afun}) explicitly determine
soft-gluon resummation up to N$^3$LL accuracy. The fourth-order coefficient 
$A^{(4)}_c$ is still unknown. Numerical approximations of $A^{(4)}_c$
\cite{Moch:2005ba} indicate that this coefficient can have a small 
quantitative effect
in practical applications of threshold resummation.
By direct inspection of Eqs.~(\ref{Acoef}) and (\ref{Dcoef}),
we note that the dependence on $c$ (the type of colliding parton)
of the perturbative functions $A_c(\as)$ and $D_c(\as)$
is entirely specified up to ${\cal O}(\as^3)$ 
by the overall colour factor $C_c$. To highlight this overall dependence,
we introduce
the notation
\begin{equation}
\label{acscaling}
A_c(\as)=C_c\left(\f{\as}{\pi}\right)
\left(1+\left(\f{\as}{\pi}\right)\gamma^{(1)}_{\rm cusp}+\left(\f{\as}{\pi}
\right)^2\gamma^{(2)}_{\rm cusp}\right)+ 
\left(\frac{\as}{\pi}\right)^4 A^{(4)}_c+{\cal O}(\as^5) \;,
\end{equation}
so that $\gamma^{(1)}_{\rm cusp}\equiv A^{(2)}_c/C_c=K/2$
and $\gamma^{(2)}_{\rm cusp}\equiv A^{(3)}_c/C_c$ (see Eq.~(\ref{Acoef}))
are universal QCD coefficients (namely, they do not depend on the type
$c$ of colliding parton). This overall dependence on $C_c$,
which is customarily named as Casimir scaling relation,
follows from the soft-parton origin of both  $A_c(\as)$ and $D_c(\as)$,
and it is eventually
a consequence
of non-abelian exponentiation \cite{Gatheral:1983cz} for soft-gluon radiation.
The validity of the Casimir scaling relation (\ref{acscaling}) beyond 
${\cal O}(\as^3)$ is a subject of current theoretical investigations (see
Ref.~\cite{Ahrens:2012qz} and references therein).
More detailed comments on the structure of soft-gluon radiation are postponed
below Eq.~(\ref{r3fin}).

In this paper we focus on the threshold-resummation factor
$C^{\, \rm th}_{c{\bar c}\to F}$.
The factor $C^{\, \rm th}_{c{\bar c}\to F}$ 
embodies all the remaining $N$-independent contributions 
(i.e., terms that are constant
in the limit $N \to \infty$) to the partonic cross section 
in Eq.~(\ref{thallorder}).
This factor is
definitely process dependent, and it has
the general perturbative expansion
\begin{equation}
C^{\, \rm th}_{c{\bar c}\to F}(\as) = 1 + \sum_{n=1}^\infty
\left(\frac{\as}{\pi}\right)^n C^{\,{\rm th}\,(n)}_{c{\bar c}\to F} \;\;.
\label{cthexp}
\end{equation}
Despite its process dependence,
in Ref.~\cite{Catani:2013tia} we have discussed and shown  
that the all-order factor $C^{\, \rm th}_{c{\bar c}\to F}(\as)$ involves a minimal
amount of process-dependent information. This information is entirely due to
the renormalized
all-loop scattering amplitude
${\cal M}_{c{\bar c}\to F}$ of the (elastic-production) partonic process
$c{\bar c}\to F$. Having ${\cal M}_{c{\bar c}\to F}$, we can introduce the
corresponding hard-virtual amplitude
$\widetilde{\cal M}^{\rm th}_{c{\bar c}\to F}$ for threshold resummation
by using a process-independent (universal) factorization formula that has the 
following all-order expression \cite{Catani:2013tia}: 
\begin{equation}
\widetilde{\cal M}^{\rm th}_{c{\bar c}\to F}
= \left[ 1 - \tilde{I}_c^{\,\rm th}(\ep,M^2) \right]  
{\cal M}_{c{\bar c}\to F}
\;\;.
\label{thvall}
\end{equation}
The subtraction operator $\tilde{I}_c^{\,\rm th}(\ep,M^2)$ in Eq.~(\ref{thvall})
is a renormalization-group invariant quantity that
does not depend on the specific final-state system $F$: it only 
depends on the type ($c=q$ or $c=g$) of colliding partons and on a scale that 
is set by the invariant mass $M$ of the system $F$. The factor 
$C^{\, \rm th}_{c{\bar c}\to F}(\as)$ is then directly related to 
$\widetilde{\cal M}^{\rm th}_{c{\bar c}\to F}$. In the simple case where
the system $F$ consists of a single particle of mass $M$, the direct relation 
is \cite{Catani:2013tia}
\begin{equation}
\label{cth}
\as^{2k}(M^2) \;C^{\,\rm th}_{c{\bar c}\to F}(\as(M^2))
=\f{|\widetilde{\cal M}^{\rm th}_{c{\bar c}\to F}|^2}{
|{\cal M}_{c{\bar c}\to F}^{(0)}|^2}\;\; ,  \quad (F: {\rm single \; particle}),
\end{equation}
where the value $k$ of the power of $\as(M^2)$ and the lowest-order amplitude
${\cal M}_{c{\bar c}\to F}^{(0)}$ are precisely defined in Eq.~(\ref{ampli}).
The relation in Eq.~(\ref{cth}) can be straightforwardly generalized to 
the more general case where the system $F$ is formed by two or more particles
with momenta $q_i$ (see Eq.~(\ref{class})). The generalization simply follows 
from the fact that we are considering the cross section integrated over the
final-state momenta $q_i$ and, therefore, we have
\begin{equation}
\label{cthmultiparticle}
\sigma^{(0)}_{c{\bar c}\to F}(M^2; \as(M^2)) \;
C^{\, \rm th}_{c{\bar c}\to F}(\as(M^2)) = \int_{PS(\{q_i\};M)} 
\;|\widetilde{\cal M}^{\rm th}_{c{\bar c}\to F}(\{q_i\})|^2 \;,\quad (F: {\rm
multiparticle \; system}).
\end{equation}
Here we have introduced a shorthand (symbolic) notation: the symbol 
$\int_{PS(\{q_i\};M)}$ denotes the properly normalized 
(see Eq.~(\ref{sigma0multiparticle})) 
phase space integration over the final-state
momenta $\{q_i\}$ at fixed value of the their total invariant mass $M$.
The extension from Eq.~(\ref{cth}) to Eq.~(\ref{cthmultiparticle})
derives from the simple 
key observation that the operator $\tilde{I}_c^{\,\rm th}(\ep,M^2)$ in
Eq.~(\ref{thvall}) is 
completely independent of the final-state
momenta $q_i$ and, therefore, the $q_i$-dependence of 
$\widetilde{\cal M}^{\rm th}_{c{\bar c}\to F}(\{q_i\})$ is entirely and directly
given by the $q_i$-dependence of the scattering amplitude 
${\cal M}_{c{\bar c}\to F}(\{q_i\})$.
In Ref.~\cite{Catani:2013tia} we obtained the explicit expression of 
the subtraction
operator $\tilde{I}_c^{\,\rm th}$ up to the second order in the 
QCD coupling $\as$.
In this paper we extend those results and compute $\tilde{I}_c^{\,\rm th}$
to the third order in $\as$.

Before presenting our results, we give more details on the notation that is used
in Eqs.~(\ref{thvall})--(\ref{cthmultiparticle}).
The all-loop scattering amplitude ${\cal M}_{c{\bar c}\to F}$
of the partonic process $c{\bar c}\to F$
contains ultraviolet (UV) and infrared (IR) singularities,
which are regularized in $d=4-2\epsilon$ space-time dimensions.
To be definite we use the customary scheme of
conventional dimensional regularization (CDR).
Before performing renormalization, the multiloop QCD amplitude 
has a perturbative dependence on
powers of $\as^u\mu_0^{2\ep}$, where $\as^u$ is the bare coupling and $\mu_0$ is
the dimensional-regularization scale. 
In the following we work with the renormalized 
on-shell scattering amplitude
that
is obtained from the corresponding unrenormalized amplitude by just expressing 
the bare coupling $\as^u$
in terms of the running coupling $\as(\mu_R^2)$ according to the \ms\ scheme
relation
\begin{equation}
\label{asren}
\as^u \,\mu_0^{2\ep}S_\ep = \as(\mu_R^2) \,\mu_R^{2\ep} \;Z(\as(\mu_R^2),\ep)
\;\;,\quad S_\ep=(4\pi)^\ep \,e^{-\ep\gamma_E} \;\;,
\end{equation}
\begin{equation}
Z(\as,\ep)=1-\as\f{\beta_0}{\ep}+\as^2
\left(\f{\beta_0^2}{\ep^2}-\f{\beta_1}{2\ep}\right)
-\as^3\left(\f{\beta_0^3}{\ep^3}-\f{7}{6}\f{\beta_0\beta_1}{\ep^2}+\f{\beta_2}{3\ep}\right)+
{\cal O}(\as^4)
\;\;,
\end{equation}
where $\gamma_E$ is the Euler number, $\mu_R$ is the renormalization scale
and $\beta_0, \beta_1$ and $\beta_2$ are the first three coefficients 
of the QCD $\beta$-function \cite{uvren}:
\begin{align}
\label{betacoeff}
&12\pi \,\beta_0=11C_A-2n_F~~,~~~~24\pi^2\,\beta_1=17C_A^2-5C_An_F-3C_Fn_F\; ,
\nn \\
&64\pi^3\beta_2=\f{2857}{54}C_A^3-\f{1415}{54}C_A^2n_F-\f{205}{18}C_AC_Fn_F+C_F^2n_F+\f{79}{54}C_An_F^2+\f{11}{9}C_Fn_F^2
\;\;.
\end{align}
The renormalized all-loop amplitude 
${\cal M}_{c{\bar c}\to F}$ has the perturbative (loop)
expansion
\begin{align}
{\cal M}_{c{\bar c}\to F} \!
&= \left( \as(M^2) \,M^{2\ep} \right)^k
\left[
{\cal M}_{c{\bar c}\to F}^{\,(0)}
\!+ \sum_{n=1}^{\infty} \left(\frac{\as(M^2)}{2\pi}\right)^{\!\!n}
\!\!{\cal M}_{c{\bar c}\to F}^{\,(n)}
\right] ,
\label{ampli}
\end{align}
where the value $k$ of the overall power of $\as$ depends on the specific
process (for instance, $k=0$ in the case of the vector boson production 
process $q{\bar q}\to V$, and $k=1$ 
in the case of the Higgs boson production process $gg \to H$
through a heavy-quark loop).
Note that the lowest-order term 
${\cal M}_{c{\bar c}\to F}^{\,(0)}$ 
is not necessarily a tree-level amplitude
(for instance, it involves a quark loop in the cases $gg \to H$ and $gg \to
\gamma \gamma$).
If $F$ is a multiparticle system, using the shorthand notation of 
Eq.~(\ref{cthmultiparticle}), we can write the lowest-order cross section as 
\begin{equation}
\label{sigma0multiparticle}
\sigma^{(0)}_{c{\bar c}\to F}(M^2; \as(M^2)) \;
= \as^{2k}(M^2) \int_{PS(\{q_i\};M)} 
\;|{\cal M}_{c{\bar c}\to F}^{\,(0)}(\{q_i\})|^2 \;\;,
\end{equation}
which (implicitly) fixes the overall normalization of the phase space 
integration. 
The perturbative terms ${\cal M}_{c{\bar c}\to F}^{\,(l)}$
$(l=1,2,3,\dots)$ are UV finite, but they still depend on $\ep$: 
in particular, they contain $\ep$-pole contributions and, therefore,
they are IR divergent as $\ep\to 0$. 
The IR divergent contributions to the scattering amplitude 
${\cal M}_{c{\bar c}\to F}$ have a universal
(process-independent) structure 
\cite{Catani:1998bh, Sterman:2002qn, Dixon:2008gr, Becher:2009qa} 
that is explicitly known up to the three-loop $(l=3)$ level \cite{Moch:2005id}.
The subtraction operator $\tilde{I}_c^{\,\rm th}(\ep,M^2)$ in Eq.~(\ref{thvall})
has the perturbative expansion
\begin{equation}
\tilde{I}_c^{\,\rm th}(\ep,M^2) = 
\sum_{n=1}^\infty \left( \frac{\as(M^2)}{2\pi} \right)^{\!\!n}
\tilde{I}_c^{\,{\rm th} (n)}(\ep) \;\;,
\label{thitilall}
\end{equation}
and the perturbative terms $\tilde{I}_c^{\,{\rm th} (n)}(\ep)$ contain IR
divergent contributions ($\ep$-poles) and a definite amount of IR finite
contributions. The IR divergent contributions to 
$\tilde{I}_c^{\,\rm th}(\ep,M^2)$ are exactly those that are necessary to cancel
the IR divergences of the renormalized all-loop amplitude 
${\cal M}_{c{\bar c}\to F}$. Therefore, the hard-virtual amplitude 
$\widetilde{\cal M}^{\rm th}_{c{\bar c}\to F}$ in Eq.~(\ref{thvall}) is IR
finite order-by-order in perturbation theory, and it can be evaluated
in the limit $\ep \to 0$. The threshold resummation
coefficient $C^{\, \rm th}_{c{\bar c}\to F}(\as(M^2))$ 
can be directly computed in the four-dimensional limit $\ep \to 0$
(though, this limit is not explicitly denoted in the right-hand side of
Eqs.~(\ref{cth}) and (\ref{cthmultiparticle})).
The perturbative expansion of $\widetilde{\cal M}^{\rm th}_{c{\bar c}\to F}$
is completely analogous to that of ${\cal M}_{c{\bar c}\to F}$ 
(see Eq.~(\ref{ampli})) with the replacement  
${\cal M}_{c{\bar c}\to F}^{\,(n)} \to 
\widetilde{\cal M}^{\rm th\,(n)}_{c{\bar c}\to F}\;$.
Note that $\widetilde{\cal M}^{\rm th\,(0)}_{c{\bar c}\to F} =
{\cal M}_{c{\bar c}\to F}^{\,(0)}\,$, and the higher-order contributions
$\widetilde{\cal M}^{\rm th\,(n)}_{c{\bar c}\to F} \;(n\geq 1)$
are obtained from Eq.~(\ref{thvall}) in terms of 
${\cal M}^{\,(l)}_{c{\bar c}\to F}$ and $\tilde{I}_c^{\,{\rm th} (l)}(\ep)$ 
at equal or lower orders, i.e. with $l \leq n$ (see, e.g., Eqs.~(48) and (49) in
Ref.~\cite{Catani:2013tia}). For simplicity,
the perturbative expansions on the right-hand side
of Eqs.~(\ref{ampli}) and (\ref{thitilall}) are expressed in powers of
$\as(M^2)$. Note, however, that ${\cal M}_{c{\bar c}\to F}$ and 
$\tilde{I}_c^{\,\rm th}(\ep,M^2)$ are separately renormalization-group invariant
quantities. Therefore, they can be equivalently expanded 
as powers series
in $\as(\mu_R^2)$, with corresponding perturbative terms that depend
on $M^2/\mu_R^2$ (see, e.g., Eqs.~(50)--(57) in Ref.~\cite{Catani:2013tia})).
The equivalent expansions are simply obtained by using Eq.~(\ref{asren}) 
to directly express $\as(M^2)$ in terms of $\as(\mu_R^2)$ and integer
powers of $(M^2/\mu_R^2)^{-\ep}$.

In Ref.~\cite{Catani:2013tia} we derived the explicit expression 
of the first-order and second-order subtraction operators
$\tilde{I}_c^{\,{\rm th} (1)}(\ep)$ and $\tilde{I}_c^{\,{\rm th} (2)}(\ep)$.
To extend the results
to the third order, we introduce a more compact (though completely equivalent)
all-order representation.
The operator $\tilde{I}_c^{\,\rm th}(\ep,M^2)$ can be written as
\begin{equation}
\label{ithexp}
1-\tilde{I}_c^{\,\rm
th}(\ep,M^2)=\exp\left\{R_c(\ep,\as(M^2))-i\Phi_c(\ep,\as(M^2))\right\}\, ,
\end{equation}
where $R_c$ and $\Phi_c$ are {\em real} functions.
The function $\Phi_c(\ep,M^2)$ is the IR divergent Coulomb phase that originates
from the virtual contributions to the all-loop amplitude 
${\cal M}_{c{\bar c}\to F}$.
Its explicit expression up to ${\cal O}(\as^3)$ \cite{Moch:2005id} reads
\begin{align}
\label{coulphase}
-i\Phi_c(\ep,\as)&=\f{i\pi\,C_c}{2\ep}\Big\{\left(\f{\as}{\pi}\right)+\left(\f{\as}{\pi}\right)^2\f{1}{2}\left(\gamma_{\rm cusp}^{(1)}-\f{\beta_0\pi}{\ep}\right) 
\nn\\
&+\left(\f{\as}{\pi}\right)^3\f{1}{3}\left(\gamma_{\rm cusp}^{(2)}-\f{1}{\ep}\gamma_{\rm cusp}^{(1)}\,\beta_0\pi+\f{1}{\ep}\pi^2\left(\f{\beta_0^2}{\ep}-\beta_1\right)\right)\Big\}+{\cal O}(\as^4)
\;\;.
\end{align}
The function $R_c(\ep ,\as)$ 
contains IR finite terms and all the remaining IR divergent terms (in the limit
$\ep\to 0$) in the exponent of Eq.~(\ref{ithexp}). This perturbative function 
can be decomposed as follows:
\begin{equation}
\label{scdecomp}
R_c(\ep,\as)=R_{c}^{\rm soft}(\ep,\as)+R_{c}^{\rm coll}(\ep,\as) \;\;,
\end{equation}
where 
\begin{equation}
\label{rsoft}
R_{c}^{\rm soft}(\ep,\as)=C_c\left(\f{\as}{\pi}R^{{\rm soft}(1)}(\ep)+
\left(\f{\as}{\pi}\right)^2 R^{{\rm soft}(2)}(\ep)+\left(\f{\as}{\pi}\right)^3
R^{{\rm soft}(3)}(\ep)\right)+{\cal O}(\as^4)\, ,
\end{equation}
\begin{equation}
\label{rcoll}
R_{c}^{\rm coll}(\ep,\as)=\f{\as}{\pi}R_c^{{\rm coll}(1)}(\ep)+
\left(\f{\as}{\pi}\right)^2 R_c^{{\rm coll}(2)}(\ep)+\left(\f{\as}{\pi}\right)^3
R_c^{{\rm coll}(3)}(\ep)+{\cal O}(\as^4)\; .
\end{equation}
The two components $R_{c}^{\rm soft}$ and $R_{c}^{\rm coll}$ of 
Eq.~(\ref{scdecomp}) have a soft and collinear origin, respectively.
The $\ep$-dependent perturbative coefficients on the right-hand side of
Eqs.~(\ref{rsoft}) and (\ref{rcoll}) read
\begin{align}
\label{soft0}
R^{{\rm soft}(1)}(\ep)&=\f{1}{2\ep^2}+R^{{\rm fin}(1)} \;,\\
\label{coll0}
R_c^{{\rm coll}(1)}(\ep)&=\f{\gamma_c}{2\ep} \;,
\end{align}
\begin{align}
\label{soft1}
R^{{\rm soft}(2)}(\ep)&=-\f{3}{8}\f{\beta_0\pi}{\ep^3}+\f{1}{8\ep^2}\gamma_{\rm
cusp}^{(1)}-\f{1}{16\ep}d_{(1)}+R^{{\rm fin}(2)} \;,\\
\label{coll1}
R_{c}^{{\rm
coll}(2)}(\ep)&=-\f{\beta_0\pi}{4\ep^2}\gamma_c+\f{1}{8\ep}\gamma_{c}^{(1)} \;,
\end{align}
\begin{align}
\label{soft2}
R^{{\rm
soft}(3)}(\ep)&=\f{11\beta_0^2-8\beta_1\,\ep}{36\ep^4}\pi^2-\f{5}{36\ep^3}\beta_0\pi\gamma^{(1)}_{\rm
cusp}+\f{1}{18\ep^2}\gamma^{(2)}_{\rm cusp}+\f{1}{24\ep^2}\beta_0\pi
d_{(1)}-\f{1}{48\ep}d_{(2)}+R^{{\rm fin}(3)} \;, \\
\label{coll2}
R_c^{{\rm
coll}(3)}(\ep)&=\f{\gamma_c}{6\ep^2}\left(\f{(\beta_0\pi)^2}{\ep}-\beta_1\pi^2\right)-\beta_0\pi\f{\gamma_c^{(1)}}{12\ep^2}+\f{1}{24\ep}\gamma_c^{(2)}\, .
\end{align}
The coefficients $\gamma_c$, $\gamma_c^{(1)}$ and $\gamma_c^{(2)}$ 
in Eqs.~(\ref{coll0}), (\ref{coll1}) and (\ref{coll2}) depend on the parton
flavour $c=q,g$ and they have a collinear origin. 
They are equal to the coefficients of the term proportional to
$\delta(1-~z)$ (i.e., to the virtual contribution) in the leading order (LO), 
next-to-leading order (NLO)  and next-to-next-to-leading order
(NNLO) collinear splitting functions 
\cite{Moch:2004pa}, and their explicit 
values\footnote{In Ref.~\cite{Catani:2013tia} we used a slightly different 
notation, and the coefficient $\gamma_{c(1)}$ therein is related to 
$\gamma_c^{(1)}$ as $\gamma_{c}^{(1)}=-\gamma_{c(1)}/8$.} are
\begin{align}
\gamma_q&=\f{3}{2}C_F \;,\nn
\end{align}
\begin{align}
\gamma_q^{(1)}&=\left(\f{3}{8}-\f{1}{2}\pi^2+6\zeta_3\right)\,C_F^2
+\left(\f{17}{24}+\frac{11}{18}\pi^2 -3\zeta_3\right)\,C_F C_A
+\left(-\frac{1}{12}-\frac{1}{9}\pi^2 \right)\,C_F n_F\nn \;,\\
\gamma_q^{(2)}&=C_F^3\left(\f{29}{16}+\f{3}{8}\pi^2+\f{\pi^4}{5}+\f{17}{2}\zeta_3-\f{2}{3}\pi^2\zeta_3-30\zeta_5\right)\nn\\
&+C_F^2C_A\left(\f{151}{32}-\f{205}{72}\pi^2-\f{247}{1080}\pi^4+\f{211}{6}\zeta_3+\f{1}{3}\pi^2\zeta_3+15\zeta_5\right)\nn\\
&+C_A^2C_F\left(-\f{1657}{288}+\f{281}{81}\pi^2-\f{\pi^4}{144}-\f{194}{9}\zeta_3+5\zeta_5\right)\nn\\
&+C_F^2n_F\left(-\f{23}{8}+\f{5}{36}\pi^2+\f{29}{540}\pi^4-\f{17}{3}\zeta_3\right)+C_Fn_F^2\left(-\f{17}{72}+\f{5}{81}\pi^2-\f{2}{9}\zeta_3\right)\nn\\
&+C_FC_An_F\left(\f{5}{2}-\f{167}{162}\pi^2+\f{\pi^4}{360}+\f{25}{9}\zeta_3\right)
\;,
\label{collq}
\end{align}
\begin{align}
\gamma_g&=\f{11}{6}C_A-\f{1}{3}n_F \;,\nn\\
\gamma_g^{(1)}&=\left(\f{8}{3}+3\zeta_3\right)C_A^2-\f{2}{3}C_A\,n_F-\f{1}{2}C_Fn_F
\;,\nn\\
\gamma_g^{(2)}&=C_A^3\left(\f{79}{16}+\f{\pi^2}{18}+\f{11}{432}\pi^4+\f{67}{3}\zeta_3-\f{1}{3}\pi^2\zeta_3-10\zeta_5\right)+C_A^2n_F\left(-\f{233}{144}-\f{\pi^2}{18}-\f{\pi^4}{216}-\f{10}{3}\zeta_3\right)\nn\\
&+\f{1}{8}C_F^2n_F-\f{241}{144}C_AC_Fn_F+\f{29}{144}C_An_F^2+\f{11}{72}C_Fn_F^2
\;.
\label{collg}
\end{align}
The coefficients $d_{(1)}$ and $d_{(2)}$ in Eqs.~(\ref{soft1}) and 
(\ref{soft2}) have a soft origin, and their values read
\begin{align}
\label{d1soft}
d_{(1)}&= \left(\frac{28}{27} - \frac{1}{18}  \pi^2\right) n_F
+ \left( -\frac{202}{27} + \frac{11}{36} \pi^2 + 7 \zeta_3
\right) C_A \;,\\
d_{(2)}&=C_A^2\left(-\f{136781}{5832} + \f{6325}{1944}\pi^2 - \f{11}{45}\pi^4+
\f{329}{6}\zeta_3 - \f{11}{9}\pi^2\zeta_3 - 24\zeta_5\right)\nn\\
&+C_A\,n_F\left(\f{5921}{2916} - \f{707}{972}\pi^2 + \f{\pi^4}{15} -
\f{91}{27}\zeta_3\right)+C_F\,n_F\left(\f{1711}{216} - \f{\pi^2}{12} - \f{\pi^4}{45} - \f{38}{9}\zeta_3\right)\nn\\
&+n_F^2\left(\f{260}{729} + \f{5}{162}\pi^2 - \f{14}{27}\zeta_3\right)\;.
\label{d2soft}
\end{align}
The coefficients $R^{{\rm fin}(1)}$ and $R^{{\rm fin}(2)}$ determine the IR finite
part on the right-hand side of Eqs.~(\ref{soft0}) and (\ref{soft1}): their
explicit values are known \cite{Catani:2013tia} and read\footnote{
In Ref.~\cite{Catani:2013tia}, the IR finite
part of $\tilde{I}_c^{\,\rm th (1)}$ and $\tilde{I}_c^{\,\rm th (2)}$
is specified by using a different notation in terms of the coefficients 
$\delta^{\,\rm th}_{(1)}$ and $\delta^{\,\rm th}_{(2)}$ therein.}
\begin{align}
R^{{\rm fin}(1)}&=-\f{\pi^2}{8} \;,\\
R^{{\rm fin}(2)}&=C_A\left(\f{607}{648}-\f{469}{1728}\pi^2+\f{\pi^4}{288}-\f{187}{144}\zeta_3\right)+n_F\left(-\f{41}{324}+\f{35}{864}\pi^2+\f{17}{72}\zeta_3\right)\, .
\end{align}
The first-order and second-order results in Eqs.~(\ref{soft0})--(\ref{coll1})
were obtained in Ref.~\cite{Catani:2013tia}. 
The three-loop expressions in Eqs.~(\ref{soft2}) and (\ref{coll2}) and,
especially, the value of the IR finite part $R^{{\rm fin}(3)}$ in Eq.~(\ref{soft2})
are the main new results of the present paper. The explicit value of the 
third-order coefficient $R^{{\rm fin}(3)}$ is
\begin{align}
R^{{\rm fin}(3)}&=\Bigg(\f{5211949}{1679616}-\f{578479}{559872}\pi^2+\f{9457}{311040}\pi^4+\f{19}{326592}\pi^6-\f{64483}{7776}\zeta_3+\f{121}{192}\pi^2\zeta_3+\f{67}{72}\zeta_3^2\nn\\
&-\f{121}{144}\zeta_5\Bigg)C_A^2+\left(-\f{412765}{839808}+\f{75155}{279936}\pi^2-\f{79}{9720}\pi^4+\f{154}{81}\zeta_3-\f{11}{288}\pi^2\zeta_3-\f{1}{24}\zeta_5\right)C_A\, n_F\nn\\
&+\left(-\f{42727}{62208}+\f{605}{6912}\pi^2+\f{19}{12960}\pi^4+\f{571}{1296}\zeta_3-\f{11}{144}\pi^2\zeta_3+\f{7}{36}\zeta_5\right)C_F\,n_F\nn\\
&+\left(-\f{2}{6561} - \f{101}{7776}\pi^2 + \f{37}{77760}\pi^4 -
\f{185}{1944}\zeta_3\right)n_F^2 \;\; .
\label{r3fin}
\end{align}

We note that the phase factor $e^{-i\Phi_c}$ in Eq.~(\ref{ithexp}) is 
physically (and practically) 
harmless to the purpose of computing the threshold resummation coefficient
$C^{\,\rm th}_{c{\bar c}\to F}$ in Eqs.~(\ref{cth}) and (\ref{cthmultiparticle}).
Indeed, $e^{-i\Phi_c}$ produces a corresponding overall phase factor 
contribution to $\widetilde{\cal M}^{\rm th}_{c{\bar c}\to F}$ in 
Eq.~(\ref{thvall}) and, therefore, $e^{-i\Phi_c}$ gives a vanishing contribution
to $|\widetilde{\cal M}^{\rm th}_{c{\bar c}\to F}|^2$ and, hence, to 
$C^{\,\rm th}_{c{\bar c}\to F}$.
We recall \cite{Catani:2013tia}
that this phase factor has been introduced in $\tilde{I}_c^{\,{\rm th}}$ to the
sole practical (aesthetical) purpose of cancelling the IR divergent Coulomb phase
of the virtual amplitude ${\cal M}_{c{\bar c}\to F}$, so that 
$\widetilde{\cal M}^{\rm th}_{c{\bar c}\to F}$ itself (and not only 
$|\widetilde{\cal M}^{\rm th}_{c{\bar c}\to F}|^2$) is IR finite in the limit
$\ep \to 0$.
We note that $\widetilde{\cal M}^{\rm th}_{c{\bar c}\to F}$ can also be redefined
by including equally harmless contributions that are purely real (rather than
phase factors). We can consider a {\em multiplicative} redefinition 
$\widetilde{\cal M}^{\rm th}_{c{\bar c}\to F} \to
F(\as,\ep) \,\widetilde{\cal M}^{\rm th}_{c{\bar c}\to F}\,$, where $F$ is an
arbitrary perturbative function (i.e., $F= 1 + {\cal O}(\as)$) such that it is
equal to unity in the limit $\ep \to 0$ (i.e., $F= 1 + {\cal O}(\ep^m)$ with
$m=1,2,\dots$). Since $\widetilde{\cal M}^{\rm th}_{c{\bar c}\to F}$ is IR
finite, this multiplicative redefinition gives a vanishing contribution
to $\widetilde{\cal M}^{\rm th}_{c{\bar c}\to F}$ in the four-dimensional limit 
$\ep \to 0$. Such harmless multiplicative redefinition corresponds to the
replacement $(1- \tilde{I}_c^{\,{\rm th}}) \to F(\as,\ep) 
\,(1- \tilde{I}_c^{\,{\rm th}})$ or, equivalently, to the replacement
$R_c(\ep,\as) \to R_c(\ep,\as) + \ln F(\as,\ep) = R_c(\ep,\as) + 
{\cal O}(\ep^m)$, with $m=1,2,\dots$, in Eq.~(\ref{ithexp})
(we have used $\ln F(\as,\ep)={\cal O}(\ep^m)$).
Therefore, we see that terms of ${\cal O}(\ep^m)$, with $m=1,2,\dots$,
in $R_c(\ep,\as)$ are harmless. In our explicit expressions 
(see Eqs.~(\ref{scdecomp})--(\ref{coll2})) of $R_c(\ep,\as)$ we have not included
any of these terms, whereas the explicit expressions of
$\tilde{I}_c^{\,{\rm th} (1)}(\ep)$ and $\tilde{I}_c^{\,{\rm th} (2)}(\ep)$
that are presented in Ref.~\cite{Catani:2013tia} include 
contributions that are due to this
type of harmless terms.

The derivation of the factorization formula (\ref{thvall}), its origin and the 
general
structure of the subtraction operator  $\tilde{I}_c^{\,{\rm th}}(\ep,M^2)$
in Eq.~(\ref{ithexp}) were discussed in Ref.~\cite{Catani:2013tia}.
Here we limit ourselves to presenting the main conclusions of our reasoning 
\cite{Catani:2013tia} in a
very concise form (we refer to Sects.~4.1 and 5 of Ref.~\cite{Catani:2013tia}
for an extended discussion).
We have already recalled the origin of the phase factor $e^{-i\Phi_c}$ 
in Eq.~(\ref{ithexp}).
We then recall 
\cite{Catani:2013tia} that the remaining contributions to 
$\tilde{I}_c^{\,{\rm th}}$ (i.e., the factor $e^{R_c}$ in Eq.~(\ref{ithexp}))
have a soft and collinear origin, as specified by the decomposition in
Eq.~(\ref{scdecomp}). The collinear contributions are embodied in the factor 
$e^{R_c^{\rm coll}}$, and they are entirely due to the {\em virtual} part
of the collinear-counterterm factor that is introduced in the (bare) partonic
cross sections to factorize the \ms\ parton densities (see Eq.~(\ref{sigtot})).
Since we are considering parton densities in the \ms\ factorization scheme,
this collinear-counterterm factor is completely and explicitly specified up to 
${\cal O}(\as^3)$ \cite{Moch:2004pa} and, in particular, the perturbative 
function $R_c^{\rm coll}(\ep,\as)$ in Eq.~(\ref{rcoll})
includes only $\ep$-pole contributions 
(see Eqs.~(\ref{coll0}), (\ref{coll1}) and (\ref{coll2})) 
with no additional IR finite terms.
The soft contributions to 
$\tilde{I}_c^{\,{\rm th}}$ are embodied in the factor 
$e^{R_c^{\rm soft}}$. They are due to the soft part of the \ms\
collinear counterterm \cite{Moch:2004pa} 
and to the inelastic processes $c {\bar c} \to F + X$,
where the radiated final-state system $X$ includes only soft partons.
The soft-parton contribution of the inelastic processes can be determined by using
universal (process-independent) soft factorization formulae 
\cite{Bassetto:1984ik, Catani:1999ss, Bern:1999ry, Catani:2000pi, Feige:2014wja} 
of the corresponding scattering amplitudes.
In Ref.~\cite{deFlorian:2012za}, the soft-parton contribution to the total cross
section was explicitly computed up to NNLO in a process-independent form by using
soft factorization formulae up to ${\cal O}(\as^2)$ 
\cite{Catani:1999ss, Bern:1999ry, Catani:2000pi}.
A corresponding process-independent calculation at N$^3$LO can be
performed by using soft factorization formulae at
${\cal O}(\as^3)$ \cite{Li:2013lsa, Li:2014bfa}.
As discussed in Ref.~\cite{Li:2014bfa}, soft-factorization results from
Refs.~\cite{Bern:1999ry, Catani:2000pi, Li:2013lsa, Li:2014bfa}
and the soft limit of the results in Refs.~\cite{real}
can be combined and used to reproduce \cite{Li:2014bfa}
the results of the N$^3$LO cross sections for Higgs boson
\cite{Anastasiou:2014vaa} and DY production \cite{Ahmed:2014cla}.
However, as discussed and pointed out in Ref.~\cite{Catani:2013tia},
much information on the soft contribution to $\tilde{I}_c^{\,{\rm th}}$ can
be obtained independently of detailed
computations.
Indeed, due to non-abelian eikonal
exponentiation \cite{Gatheral:1983cz}, the intensity of soft radiation from
the parton $c$ is simply proportional to the Casimir coefficient $C_c$ of that
parton (this conclusion is certainly valid up to ${\cal O}(\as^3)$
\cite{Gatheral:1983cz}). Therefore, $R_{c}^{\rm soft}(\ep,\as)$ can be expressed
by factorizing the overall coefficient $C_c$ as in Eq.~(\ref{rsoft}).
This Casimir scaling behaviour is completely analogous to that of the functions
$A_c(\as)$ (see Eq.~(\ref{acscaling})),
$D_c(\as)$ (see Eqs.~(\ref{dfun}) and (\ref{Dcoef}))
and $\Phi_c(\ep,\as)$ (see Eq.~(\ref{coulphase})),
since all these functions are entirely due to soft-parton contributions
\cite{Catani:2013tia}.
The perturbative coefficients $R^{{\rm soft}(n)}(\ep)$, with $n=1,2,3$, in
Eq.~(\ref{rsoft}) are completely process independent and they can be determined by
considering a single specific process. 
In particular, $R^{{\rm soft}(n)}(\ep)$ contains IR divergent contributions 
($\ep$-pole terms) and IR finite contributions. These IR divergent terms of
soft-parton origin are due to real emission contributions, but they are
constrained (because of the real--virtual cancellation mechanism of IR
divergences) to be exactly equal to the corresponding IR divergent terms
due to virtual radiation. Therefore,
the $\ep$-pole terms in 
Eqs.~(\ref{soft0}), (\ref{soft1}) and (\ref{soft2}) are completely specified by the
explicit calculation of either the quark or gluon form factors
\cite{Moch:2005id} (as recalled below, 
the process independence of these terms is consistent with the
universality structure of the IR divergent contributions to the QCD scattering
amplitudes
\cite{Catani:1998bh, Dixon:2008gr, Becher:2009qa}).
It follows that
the IR finite coefficients $R^{{\rm fin}(n)}$ ($n=1,2,3$) are the only
terms that are not explicitly determined by using our general reasoning
\cite{Catani:2013tia}. Owing to their universality, the explicit computation of a
single process is sufficient to extract the values of these IR finite coefficients. As
illustrated below,
we use the N$^3$LO Higgs boson results of Ref.~\cite{Anastasiou:2014vaa}
to obtain the value of $R^{{\rm fin}(3)}$ in Eq.~(\ref{r3fin}).

Before considering the evaluation of $R^{{\rm fin}(3)}$, we present some
additional comments on the structure of Eqs.~(\ref{ithexp})--(\ref{d2soft})
and on the connection between real- and virtual-emission contributions.
As we have discussed, the subtraction operator $(1- \tilde{I}_c^{\,{\rm th}})$
in Eqs.~(\ref{thvall}) and (\ref{ithexp}) includes the Coulomb phase factor
$e^{-i\Phi_c}$ and an additional factor of soft and collinear origin.
In Eq.~(\ref{ithexp}) we express this additional factor by using the
exponentiated form $e^{R_c}$. The exponentiated form, which is completely
equivalent to its direct expansion in powers of $\as$, is more compact in view
of the factorization and exponentiation properties of both soft and collinear
contributions. Owing to factorization we can write  
$e^{R_c}= e^{R_c^{\rm coll}} e^{R_c^{\rm soft}}$, i.e. we can introduce the
decomposition in Eq.~(\ref{scdecomp}).
The collinear factor $e^{R_c^{\rm coll}}$ is entirely due to the virtual part
of the collinear counterterm of the \ms\ parton densities, and its exponentiated
structure is eventually a consequence of the customary solution of the
Altarelli--Parisi evolution equations in terms of an exponentiated evolution
operator. Indeed (as stated below Eq.~(\ref{coll2})) the exponent 
$R_c^{\rm coll}$ is directly determined by the coefficients $\gamma_c, 
\gamma_c^{(1)}$ and $\gamma_c^{(2)}$ of the virtual part of the 
Altarelli--Parisi splitting functions.
The factor $e^{R_c^{\rm soft}}$ is due to real emission of soft partons: it
fulfils non-abelian eikonal exponentiation and, therefore, we can express the
exponent $R_c^{\rm soft}$ through the Casimir scaling relation (\ref{rsoft}).
The soft/collinear structure of $(1- \tilde{I}_c^{\,{\rm th}}) \propto
e^{R_c^{\rm coll}} e^{R_c^{\rm soft}}$ does not originate from virtual
contributions to the scattering amplitude ${\cal M}_{c{\bar c}\to F}$,
but the IR divergent terms in Eq.~(\ref{rsoft})--(\ref{coll2})
exactly match the analogous universal structure of the IR divergent virtual
contributions to ${\cal M}_{c{\bar c}\to F}$.
The IR divergent virtual contributions 
\cite{Catani:1998bh, Sterman:2002qn, Dixon:2008gr, Becher:2009qa, Moch:2005id}
include dominant and subdominant $\ep$-poles. The dominant poles have a
soft--collinear origin and are controlled by the perturbative function
$A_c(\as)$ in Eq.~(\ref{afun}) or, equivalently, the function
$\gamma_{\rm cusp}(\as)$ in Eq.~(\ref{acscaling}). The subdominant poles
originate from either collinear (and non-soft) or soft (and non-collinear)
contributions and they are controlled by the collinear coefficients in
Eqs.~(\ref{collq})--(\ref{collg}) and the soft coefficients in 
Eqs.~(\ref{d1soft})--(\ref{d2soft}).
We also note that the real emission contribution to the partonic cross section
of Eq.~(\ref{thallorder})
is separated in two different factors: the $N$-independent factor 
$e^{R_c^{\rm soft}}$ (which contributes to 
$(1- \tilde{I}_c^{\,{\rm th}})$ and, hence, to 
$C^{\, \rm th}_{c{\bar c}\to F}$) and the $\ln N$-dependent radiative factor
$\Delta_{c, \,N}$ of Eq.~(\ref{delformfact}).
These two factors have a soft origin and they are not fully independent. In
particular, the coefficients of the dominant IR poles of
${R_c^{\rm soft}}(\ep, \as)$ are directly related to the dominant 
$\ln N$-dependence of $\Delta_{c, \,N}$ (as given by the perturbative function
$A_c(\as)$). The subdominant $\ln N$-dependence of $\Delta_{c, \,N}$
is due to the soft-parton function $D_c(\as)$, whose perturbative coefficients
$D_c^{(n)}$ are related to the soft-parton coefficients $C_c \,d_{(n-1)}$ and
$C_c \,R^{{\rm fin}(n-1)}$ of ${R_c^{\rm soft}}(\ep, \as)$: this relation between 
$\ln N$ terms, $\ep$-poles and IR finite terms is discussed and worked out
in Refs.~\cite{Moch:2005ky, Laenen:2005uz}.
We note that using the general analysis of 
Refs.~\cite{Moch:2005ky, Laenen:2005uz}
and our result for 
$R^{{\rm fin}(3)}$ in Eq.~(\ref{r3fin}), the fourth-order coefficient
$D_c^{(4)}$ of $D_c(\as)$ can be determined in terms of the $\ep$-poles
at ${\cal O}(\as^4)$ (once they become available).

To evaluate the third-order coefficient $R^{{\rm fin}(3)}$, we consider the
perturbative expansion of the resummation formula in
Eq.~(\ref{thallorder}), which contains all the terms which are not suppressed in the
large-$N$ limit, namely, the logarithmically-enhanced terms and the constant terms as $N\to\infty$. We consider the N$^3$LO contribution
(see, e.g., the Appendix~E in Ref.~\cite{Catani:2003zt}) 
and we transform it back to $z$ space to obtain the general expression
of the N$^3$LO term $g_{c{\bar c}}^{F(3)}(z)$ of Eq.~(\ref{sigmahat})
in the threshold limit $z \to 1$. We find
\begin{align}
g_{c{\bar c}}^{F(3)}(z)&=8\left(A_c^{(1)}\right)^3 {\cal D}_5-\f{40}{3}\beta_0\pi\left(A_c^{(1)}\right)^2\, {\cal D}_4\nn\\
&+\left(-\f{32}{3}\pi^2 \left(A_c^{(1)}\right)^3+8C_{c {\bar c}\to F}^{{\rm th}(1)}\left(A_c^{(1)}\right)^2+16 A_c^{(1)}A_c^{(2)}+\f{16}{3}(\beta_0\pi)^2\, A_c^{(1)}\right){\cal D}_3\nn\\
&+\Bigg(160\zeta_3\left(A_c^{(1)}\right)^3-4\beta_0\pi\,A_c^{(1)}C_{c {\bar c}\to F}^{{\rm th}(1)}+8\beta_0\pi^3\,\left(A_c^{(1)}\right)^2-8\beta_0\pi\, A_c^{(2)}\nn\\
&+6A_c^{(1)}D_c^{(2)}-4A_c^{(1)}\beta_1\pi^2\Bigg){\cal D}_2\nn\\
&+\Bigg(4\left(A_c^{(3)}+A_c^{(2)}C_{c {\bar c}\to F}^{{\rm th}(1)}+A_c^{(1)}C_{c {\bar c}\to F}^{{\rm th}(2)}\right)-\f{16}{3}A^{(1)}_c\, A_c^{(2)}\pi^2\nn\\
&-\f{8}{3}\left(A_c^{(1)}\right)^2\, C_{c {\bar c}\to F}^{{\rm th}(1)}\pi^2-\f{4}{9}\pi^4 \left(A_c^{(1)}\right)^3-4\beta_0\,\pi\left(D_c^{(2)}+24\left(A_c^{(1)}\right)^2\zeta_3\right)\Bigg){\cal D}_1\nn\\
&+\Bigg(\left(192\zeta_5-\f{64}{3}\pi^2\zeta_3\right)\left(A_c^{(1)}\right)^3
+16A_c^{(1)}\zeta_3\left(2A_c^{(2)}+A_c^{(1)}C_{c {\bar c}\to F}^{{\rm th}(1)}\right)+\f{4}{9}\left(A_c^{(1)}\right)^2\beta_0\pi^5\nn\\
&+C_{c {\bar c}\to F}^{{\rm th}(1)}D_c^{(2)}+D_c^{(3)}-\f{2}{3}A_c^{(1)}D_c^{(2)}\pi^2\Bigg){\cal D}_0\nn\\
&+\Bigg(C_{c {\bar c}\to F}^{{\rm th}(3)}-\f{2}{45}A_c^{(1)}A_c^{(2)}\pi^4-\f{1}{45}\left(A_c^{(1)}\right)^2C_{c {\bar c}\to F}^{{\rm th}(1)}\pi^4
+\left(\f{160}{3}\zeta_3^2-\f{116}{2835}\pi^6\right)\left(A_c^{(1)}\right)^3\nn\\
&+4A_c^{(1)}D_c^{(2)}\zeta_3+\f{16}{3}\left(A_c^{(1)}\right)^2\beta_0\pi\left(\pi^2 \zeta_3-12\zeta_5\right)\Bigg)\delta(1-z)
+ \dots \;\;,
\label{g3sv}
\end{align}
where ${\cal D}_m = {\cal D}_m(z)$ are the plus-distributions defined in
Eq.~(\ref{plusdist}), and the dots in the right-hand side of Eq.~(\ref{g3sv})
denote additional terms that are less singular in the limit $z \to 1$
(i.e., terms that are relatively suppressed by some powers of $(1-z)$).
The terms that are explicitly denoted in the right-hand side of Eq.~(\ref{g3sv}) 
define the soft-virtual (SV) approximation of the N$^3$LO contribution 
$g_{c{\bar c}}^{F(3)}(z)$ to the partonic cross section. 
These terms depend on the universal perturbative coefficients 
$A_c^{(n)}, D_c^{(n)}$ (see Eqs.~(\ref{Acoef}) and (\ref{Dcoef}))
and on the process-dependent coefficients
$C_{c {\bar c}\to F}^{{\rm th}(n)}$ with $n \leq 3$.

In the case of Higgs boson production ($gg\to H$) by gluon fusion,
the SV N$^3$LO 
expression in Eq.~(\ref{g3sv}) exactly corresponds
to the result of the explicit computation performed in
Ref.~\cite{Anastasiou:2014vaa}.
The first-order and second-order coefficients 
$C_{gg \to F}^{{\rm th}(1)}$ and $C_{gg \to F}^{{\rm th}(2)}$
are known (they can be determined by our process-independent resummation formalism
up to ${\cal O}(\as^2)$ or, equivalently, they can be extracted from 
the SV NNLO results of Refs.~\cite{Catani:2001ic, Harlander:2001is}).
Therefore, comparing Eq.~(\ref{g3sv}) with the result in Eq.~(10) of 
Ref.~\cite{Anastasiou:2014vaa}, we can extract the coefficient 
$C_{gg \to F}^{{\rm th}(3)}$ and we find
\begin{align}
C^{{\rm th}(3)}_{gg\to H}&=C_A^3\left(\f{215131}{5184} + \f{16151}{7776}\pi^2 - \f{1765}{15552}\pi^4 + \f{1}{2160}\pi^6 - 
 \f{15649}{432}\zeta_3 - \f{77}{144}\pi^2\zeta_3 + \f{3}{2}\zeta_3^2 + 
 \f{869}{144}\zeta_5\right)\nn\\
&+C_A^2n_F\left(-\f{98059}{5184}-\f{35}{243}\pi^2+\f{2149}{38880}\pi^4+\f{29}{8}\zeta_3-\f{29}{72}\pi^2\zeta_3+\f{101}{72}\zeta_5\right)\nn\\
&+C_AC_F\,n_F\left(-\f{63991}{5184} - \f{71}{216}\pi^2 + \f{11}{6480}\pi^4 + \f{13}{2}\zeta_3 + 
\f{1}{2}\pi^2\zeta_3 +\f{5}{2}\zeta_5\right)\nn\\
&+C_F^2\,n_F\left(\f{19}{18}+\f{37}{12}\zeta_3-5\zeta_5\right)+C_A n_F^2\left(\f{2515}{1728} - \f{133}{1944}\pi^2 - \f{19}{3240}\pi^4 + \f{43}{108}\zeta_3\right)\nn\\
&+C_F\, n_F^2\left(\f{4481}{2592}-\f{23}{432}\pi^2-\f{1}{3240}\pi^4-\f{7}{6}\zeta_3\right)\, .
\label{c3H}
\end{align}
To be precise, the coefficient $C_{gg \to H}^{{\rm th}(3)}$ in Eq.~(\ref{c3H})
corresponds to the perturbative expansion that is defined by Eq.~(\ref{sigmahat})
after having rescaled the partonic cross section with the Wilson coefficient 
of the effective point-like coupling $ggH$ \cite{Chetyrkin:1997un} 
(this definition exactly corresponds to that used in Eq.~(4) of 
Ref.~\cite{Anastasiou:2014vaa}).
Having the information in Eq.~(\ref{c3H}) and using Eqs.~(\ref{thvall}) and
(\ref{cth}), we apply the operator $(1-\tilde{I}_c^{\,\rm th})$ of Eq.~(\ref{ithexp})
to the three-loop gluon form factor \cite{qgformfactor} and we can extract 
the coefficient $R^{{\rm fin}(3)}$ in Eq.~(\ref{soft2}). We find the explicit value
that is presented in Eq.~(\ref{r3fin}).

The coefficient $R^{{\rm fin}(3)}$ completely determines the explicit expression
of the process-independent subtraction operator $\tilde{I}_c^{\,\rm th}$
up to ${\cal O}(\as^3)$. Using this expression and 
Eqs.~(\ref{thvall})--(\ref{cthmultiparticle}), the threshold resummation
coefficient
$C_{c{\bar c}\to F}^{{\rm th}}(\as)$ for an arbitrary process $c{\bar c}\to F$
is straightforwardly and explicitly computable up to the three-loop order once 
the corresponding three-loop scattering amplitude ${\cal M}_{c{\bar c}\to F}$ 
for that process is known.

As an application of our general formalism and results, 
we can consider the production
of a vector boson $V$ ($V=Z,W^\pm$) by the DY process
$q{\bar q}\to V$.
Using the subtraction operator $(1-{\tilde I}_c^{\,\rm th})$ and 
the results for the quark form factor up to three-loop order \cite{qgformfactor}, 
we can compute the coefficients $C^{{\rm th}(n)}_{q{\bar q}\to V}$ with $n=1,2,3$. 
We find
\begin{equation}
C^{{\rm th}(1)}_{q{\bar q}\to V}=C_F\left(-4+\f{\pi^2}{3}\right) \;,
\label{c1dy}
\end{equation}
\begin{align}
C^{{\rm th}(2)}_{q{\bar q}\to V}&=C_F^2\left(\f{511}{64}-\f{35}{48}\pi^2+\f{\pi^4}{40}-\f{15}{4}\zeta_3\right)\nn\\
&+C_FC_A\left(-\f{1535}{192}+\f{37}{54}\pi^2-\f{\pi^4}{240}+\f{7}{4}\zeta_3\right)+C_Fn_F\left(\f{127}{96}-\f{7}{54}\pi^2+\f{1}{2}\zeta_3\right)
\;,
\end{align}
\begin{align}
C^{{\rm th}(3)}_{q{\bar q}\to V}&=C_F^3\left(-\f{5599}{384}-\f{65}{576}\pi^2-\f{17}{320}\pi^4+\f{803}{136080}\pi^6-\f{115}{16}\zeta_3+\f{5}{24}\pi^2\zeta_3+\f{1}{2}\zeta_3^2+\f{83}{4}\zeta_5\right)\nn\\
&+C_F^2\,C_A\left(\f{74321}{2304} - \f{6593}{5184}\pi^2 + \f{94}{1215}\pi^4 - \f{2309}{272160}\pi^6 - \f{8653}{432}\zeta_3 + \f{53}{54}\pi^2\zeta_3 + \f{37}{12}\zeta_3^2- \f{689}{72}\zeta_5\right)\nn\\
&+C_A^2\,C_F\Big(-\f{1505881}{62208} + \f{281}{128}\pi^2 + \f{14611}{311040}\pi^4 + \f{829}{272160}\pi^6 + 
 \f{82385}{5184}\zeta_3 - \f{221}{288}\pi^2\zeta_3\nn\\
&- \f{25}{12}\zeta_3^2 - 
 \f{51}{16}\zeta_5\Big)+C_AC_Fn_F\left(\f{110651}{15552} - \f{7033}{7776}\pi^2- \f{1439}{77760}\pi^4 - \f{94}{81}\zeta_3 + \f{13}{72}\pi^2\zeta_3 - \f{\zeta_5}{8}\right)\nn\\
&+C_F^2\,n_F\left(-\f{421}{192} + \f{329}{1296}\pi^2 - \f{223}{19440}\pi^4 + \f{869}{216}\zeta_3 - \f{7}{27}\pi^2\zeta_3 - \f{19}{18}\zeta_5\right)\nn\\
&+C_F\,n_F^2\left(-\f{7081}{15552}+\f{151}{1944}\pi^2+\f{\pi^4}{486}-\f{79}{324}\zeta_3\right)\nn\\
&+C_FN_{F,V}\left(\f{N_c^2-4}{N_c}\right)\left(\f{1}{8}+\f{5}{96}\pi^2-\f{\pi^4}{2880}+\f{7}{48}\zeta_3-\f{5}{6}\zeta_5\right)
\;,
\label{c3dy}
\end{align}
where $N_{F,V}$ is a factor originating by diagrams where the virtual
gauge boson does not couple directly to the initial state quarks \cite{qgformfactor},
and it is proportional to the charge weighted sum of the quark flavours.
The explicit expressions of the coefficients 
$A_c^{(n)}$ and $D_c^{(n)}$ up to ${\cal O}(\as^3)$ and the expressions
of  $C^{{\rm th}(n)}_{q{\bar q}\to V}$ in Eqs.~(\ref{c1dy})--(\ref{c3dy})
can be inserted in Eq.~(\ref{g3sv}) to obtain the explicit expression of the
SV N$^3$LO cross section for
the DY process. The ensuing result is in agreement with the result
in Ref.~\cite{Ahmed:2014cla}.

In this paper we have considered the processes in which an arbitrary colourless 
system $F$ with high mass is produced in hadronic collisions.
We have focused on the structure of the perturbative QCD contributions near
partonic threshold. Such contributions are controlled by universal
resummation factors plus a process dependent
hard-virtual function. As discussed in Ref.~\cite{Catani:2013tia},
the hard-virtual function is directly related to the process-dependent
virtual amplitude through a universal 
factorization formula that depends on a process-independent
subtraction operator.
The results that were documented in 
Ref.~\cite{Catani:2013tia} determine
the structure of the subtraction operator (and, thus, of 
the hard-virtual function) up to a universal perturbative function with purely
numerical perturbative coefficients that were explicitly computed 
up to the second-order in $\as$.
In this paper we have pointed out that the recent computation of 
the soft-virtual
corrections to Higgs boson production at N$^3$LO \cite{Anastasiou:2014vaa} 
is sufficient to extend those results to the third-order in $\as$, and we have
explicitly computed the corresponding perturbative coefficient.
The results presented in this paper can be used to perform soft-gluon
resummation up to N$^3$LL accuracy\footnote{A quantitative study of Higgs 
boson production at N$^3$LL accuracy, with the inclusion of the 
soft-virtual contribution at N$^3$LO, is presented in a very recent paper 
\cite{Bonvini:2014joa}.}
for the production of an arbitrary 
colourless system $F$ in hadron collisions.
Equivalently,
they allow us to determine the explicit form of the N$^3$LO corrections 
to the production cross section near
partonic threshold, once the corresponding three-loop scattering amplitude
${\cal M}_{c{\bar c}\to F}$ is available.
We have applied our results to the DY process and we have presented 
the explicit
expression of the hard-virtual function up to N$^3$LO, confirming the
result of Ref.~\cite{Ahmed:2014cla} for the DY cross section at
N$^3$LO.

\noindent {\bf Acknowledgements.}

We would like to thank Thomas Gehrmann for comments on the manuscript.
This research was supported in part by the Swiss National Science Foundation (SNF) under contract 200021-144352 and by 
the Research Executive Agency (REA) of the European Union under the Grant Agreements PITN--GA--2010--264564 ({\it LHCPhenoNet}) and PITN--GA--2012--316704 ({\it Higgstools}).

\end{document}